# Highly reproducible, large scale inkjet-printed Ag nanoparticles-ink SERS substrate

Samir Kumar[a,*], Kyoko Namura[a], Daisuke Kumaki[b], Shizuo Tokito[b], Motofumi Suzuki[a]

[a]Department of Micro Engineering, Graduate School of Engineering, Kyoto University, Katsura, Nishikyo, Kyoto 615-8540 Japan.
[b]Research Center for Organic Electronics, Yamagata University, 4-3-16, Jonan, Yonezawa, Yamagata, 992−8510, Japan.

**Abstract**

We report the fabrication of a low-cost, and highly reproducible large scale surface-enhanced Raman spectroscopy (SERS) substrate using an inkjet-printed Ag nanoparticle-ink (AgNI). The AgNI SERS substrates were evaluated for SERS using trans-1,2-bis(4-pyridyl)ethylene (BPY) as a molecular probe. AgNI substrates significantly enhanced the SERS signals with the limit of detection of $1.8 \times 10^{-7}$ g/mL for BPY. The printed AgNI dot arrays exhibit an excellent SERS performance and reproducibility. The batch-to-batch and spot-to-spot standard deviation value of less than 10% was obtained. The results reveal the reproducibility of the AgNI SERS dot arrays and its potential application for SERS substrates.

# 1. Introduction

Surface-enhanced Raman scattering (SERS) is the enormous enhancement of the Raman signal of the scattering molecules adsorbed onto metal surfaces under certain conditions.[1] The localized surface plasmon resonances (LSPR) of the metal nanoparticles of the substrate is predominantly responsible for the amplification of the local electromagnetic fields that allow for the detection of analytes at exceptionally low concentrations, even at a single molecular level.[2–4] An ideal SERS substrate should be able to generate an intense, reproducible,

*Corresponding author.
  *Email address:* samiratwork@gmail.com

and uniform SERS signal.[5–7] It should also be inexpensive and easy to fabricate on a large scale.[8]

There are various studies on the large-scale fabrication of SERS substrates fabricated by different methods. These methods include electrochemical,[9] lithography,[10] dyeing,[11] electrospinning,[12] inkjet printing,[13] and screen printing[14]. Presently, all the methods described above for SERS substrates are often complicated and laborious. Wu et al. fabricated SERS substrates utilizing screen printing Ag nanoparticles on a plastic PET (Polyethylene terephthalate) substrates.[14] Yu et al. fabricated a SERS substrate on paper by inkjet printing.[13] With most of the fabrication method, the standard deviation (SD) value of the SERS signal is between 10 – 20%. The poor reproducibility of the SERS substrate is one of the major challenges for the development of SERS sensing technology.

Inkjet printing is suitable for the large-area fabrication of SERS substrates where patterns can be formed in a single step. It is a simple fabrication process in which the ink is used for the printing of plasmonic patterns.

Therefore, many efforts have been devoted to developing highly reproducible SERS substrates. Herein, we report a large-scale fabrication method of flexible, and highly reproducible Ag nano-ink (AgNI) SERS dot arrays substrates by employing an inkjet-printing method. The printed AgNI dot arrays exhibit an excellent SERS performance and reproducibility. A batch-to-batch and spot-to-spot SD value of less than 10% has been obtained. The results demonstrate the reproducibility of the AgNI SERS dot arrays and its potential application in the field of trace detection.

## 2. Experimental details

*2.1. Fabrication of Ag nano ink printed SERS substrates*

Ag nanoparticles (AgNPs) were synthesized using oxalate-bridged silver amine complexes.[15,16] The typical particle sizes were 10 to 15 nm (d50). The AgNPs were dissolved in the mixture solvent of dodecane and 1-nonanol with 35 wt%. The prepared AgNI was patterned by an inkjet printer (Fujifilm:DMP-2831) on Polyethylene naphthalate (PEN) film with a 125 μm thickness. The jetting condition of AgNI was optimized by controlling a waveform of applying voltage in cartridge. The AgNI droplets were deposited with a dot-to-dot spacing of 60 mm. The Ag dot patterns of 5 mm diameter were formed on the PEN film. After inkjet pattering, the PEN film was baked on the hotplate for 30 min at 150ºC. The AgNPs are fused with each other by the sintering process, and then, conductive passes are formed in Ag layer. The thickness of the Ag layer was about 100 nm.

*2.2. Optical and SERS measurements*

The optical property of the AgNI dot arrays substrates were investigated using UV-Vis reflectance measurements. The reflectance spectra were collected using an Ocean Optics (USB4000) spectrometer. LambdaVision (RAM200, ) Raman spectrometer was used for the collection of SERS spectra. trans-1,2-bis(4-pyridyl) ethylene (BPY, 99.9+%, Sigma) was used as the molecular probe for the SERS study. A 785 nm laser with a 50× objective and 15 mW power on the sample was used for the excitation with a spectral acquisition time of 1 s. The SERS measurements were performed at seven random spots on the substrate. All the measurements were performed at room temperature. A 10 μL droplet of BPY solution was released onto the AgNI substrates and allowed to dry prior to data acquisition.

*2.3. Surface morphology*

A scanning electron microscope (SEM; Hitachi High Tech. SU3800) with a $LaB_6$ detector in the secondary electron mode operating at an acceleration voltage of 10 kV was used to conduct surface morphology analysis of the baked and unbaked samples.

**3. Results and discussion**

*3.1. Morphology of AgNI dot arrays*

Figure 1(d) shows the photograph of the printed unbaked and baked AgNI dot arrays on the PEN substrate. The AgNPs stick firmly to the PEN substrate and form a uniform Ag layer. The SEM micrographs of the baked and unbaked AgNI dot arrays are shown in Fig. 1. The Fig. 1(a) and 1(b) indicate that for unbaked AgNI most of the AgNPs were quasi-spherical with a mean diameter of 20 ± 5 nm. The unbaked AgNI samples were uniformly distributed. Whereas after baking the AgNPs coalesce with the formation of percolation holes. The uniform surface coverage is a crucial factor that determines the spot-spot SERS uniformity.

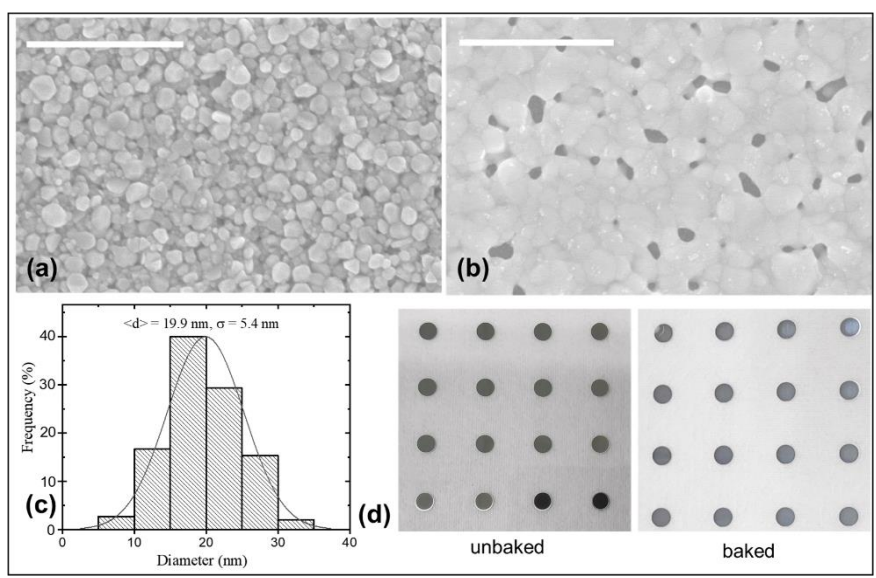

*Figure 1. SEM micrographs of (a) unbaked, and; (b) baked AgNI dot arrays samples. Scale bar equals to 500 nm; (c) particle size distribution on unbaked AgNI sample. <d> denotes the mean diameter, and σ is the standard deviation.; (d) A photograph of unbaked and baked AgNI samples.*

### 3.2. Optical and SERS characterization

The UV-Vis reflectance spectroscopy shows a reflectance minima at around 420 nm, which is due to LSPR of AgNPs. The unbaked AgNI sample showed higher absorption around the LSPR wavelength and lower reflectance at long wavelengths suggesting a better LSPR performance. The overall reflectance of the AgNI increases after the baking process, and that maybe because of the more uniform distribution of the Ag. The SERS spectrum of the AgNI substrates was taken by dropping 10 μL of $10^{-4}$ aqueous solution of BPY on the substrate. The SERS spectra of the bare PEN substrate, BPY on baked and unbaked AgNI samples are shown in Fig. 2(b). The PEN substrate shows sharp peaks at around 780, 1222, 1390, 1480, 1635, and 1719 cm$^{-1}$. Upon printing of the AgNI dot arrays, the background peaks of the PEN substrate are suppressed, and the characteristic BPY peaks can be observed. The SERS spectra of BPY has four main bands at around 1013, 1228, 1294, and 1610 cm$^{−1}$ that are associated to the pyridine ring breathing, ring deformation, the C = C in-plane ring mode, and the C = C stretching mode, respectively.[17,18] The unbaked AgNI sample showed a better SERS response as compared to that of the baked AgNI. We were able to detect BPY down to a concentration of 1.8 × 10$^{-7}$ g/mL on the unbaked AgNI dot arrays showing a good sensing capability of the unbaked AgNI sample. The better SERS response of the unbaked AgNI is because AgNPs had a quasi-spherical shape with a high degree of uniformity. The SERS intensity of Raman probe is dependent on the size and shape of the nanoparticles on whicjh they are adsorbed.[19] The uniform quasi-spherical AgNPs may enhance the SERS signal not only due to the LSPR. They may also produce unusually large electromagnetic enhancement around the sharp edges, referred to as a lightning rod effect.[20] The electric field that is generated at the sharp edges would be much greater than other parts of the nanoparticles. The plasmonic coupling of the LSPR between two AgNPs with a separation of less than 50 nm may also further enhance the SERS intensity.[21]

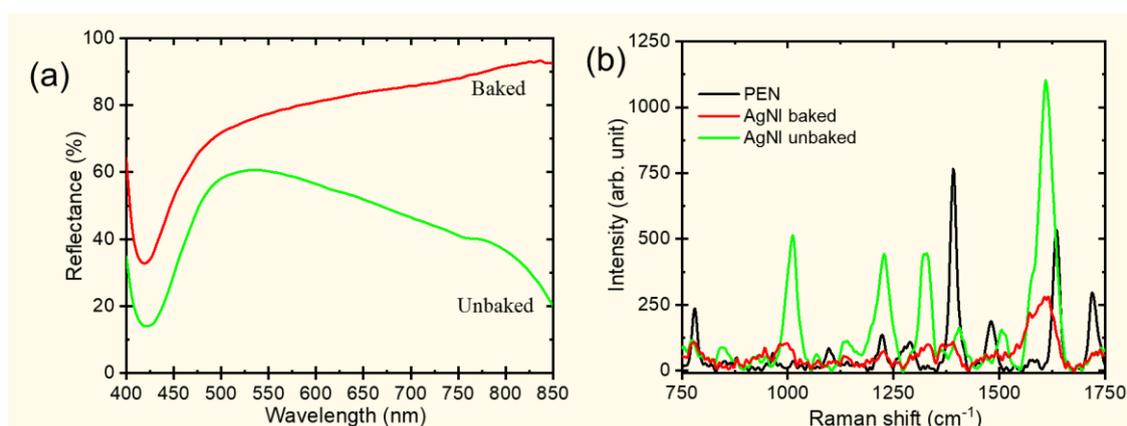

*Figure 2. (a) UV-Vis reflectance spectra for baked and unbaked AgNI samples. Aa reflectance minimum at around 420 nm, which is due to LSPR of the Ag nanoparticles; (b) Raman spectrum of the bare PEN substrate and SERS spectra of BPY on unbaked and baked AgNI dot arrays.*

### 3.3. Reproducibility of AgNI SERS substrates

Batch-batch and spot-spot uniformity on the same sample are critical concerns for large scale SERS substrates fabrication. A common barrier to quantitative detection for the SERS substrates is spot-to-spot reproducibility of the signal.[22] The intensity peak at 1610 cm$^{-1}$ for 10$^{-4}$ M BPY solution was used to measure the batch-batch and spot-spot SERS signal reproducibility, Fig. 3. Batch-batch variation of the SERS signal was very low, with a standard deviation of less than 10%. The standard deviation of the SERS measurements at different positions on the sample was also around 6%. This shows a good reproducibility of our printed AgNI SERS samples. The reproducibility is related to the uniform distribution of AgNS hot spots.

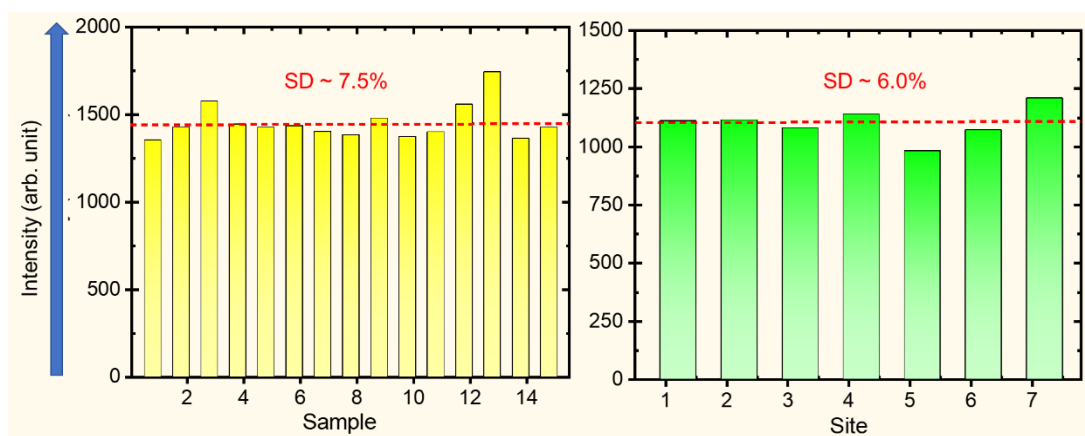

Figure 3. (a) Batch-to-batch variation of the peak around 1610 cm$^{-1}$ peak; (b) spot-to-spot variation and the standard deviation of the SERS measurements at different positions on the AgNI sample.

## 4. Conclusions

In summary, we fabricated a large scale highly reproducible inkjet-printed Ag nano ink SERS substrate. The minimum detection limit was 1.8 × 10$^{-7}$ g/mL for BPY molecules. The inkjet-printed AgNI sample displayed excellent reproducibility with batch-to-batch and spot-to-spot standard deviation of less than 10%. This study can result in large-scale production of flexible SERS substrates.

## 5. Competing interests

The authors declare that they have no competing interests.

## 6. Funding

This work was supported by JST COI Grant Number JPMJCE.1307 and JPMJCE.1312

*May 20, 2020*

# Acknowledgments

The authors also thank Dr. Kosuke Ishikawa of Kyoto University for assisting us with the SEM observations.